\documentclass[aps,prc,preprint,superscriptaddress]{revtex4-1}
\usepackage{graphicx}
\begin{document}

%\preprint{}

\title{Enhancement of fusion rates due to quantum effects in the particles momentum distribution in nonideal media}


%%%%%%%%%%%%%%%%%%%%%%%%%%%%% AUTHORS %%%%%%%%%%%%%%%%%%%%%%%%%%%%%%%%%%%%%%%%%%%%%%%%%%%%%%%%%%%
%
\author{N. J. Fisch}
%\thanks{}
\affiliation{Department of Astrophysical Sciences, Princeton University, Princeton, N.J. 08540, USA}

\author{M. G. Gladush}
\affiliation{SRC RF Troitsk Institute for Innovation and Fusion Research, Troitsk, Moscow region, 142190 Russia}

\author{Yu. V. Petrushevich}
\affiliation{SRC RF Troitsk Institute for Innovation and Fusion Research, Troitsk, Moscow region, 142190 Russia}

\author{Piero Quarati}
\affiliation{Politecnico di Torino Department of Physics, Torino I-10125, Italy and INFN, Sezione di Cagliari, Italy}

\author{A. N. Starostin}
\affiliation{SRC RF Troitsk Institute for Innovation and Fusion Research, Troitsk, Moscow region, 142190 Russia}

\date{\today}

\begin{abstract}
This study concerns a situation when measurements of the nonresonant cross-section of nuclear reactions appear highly dependent on the environment in which the particles interact. An appealing example discussed in the paper is the interaction of a deuteron beam with a target of deuterated metal Ta. In these experiments, the reaction cross section for d(d,p)t was shown to be orders of magnitude greater than what the conventional model predicts for the low-energy particles. In this paper we take into account the influence of quantum effects due to the Heisenberg uncertainty principle for particles in a non-ideal medium elastically interacting with the medium particles. In order to calculate the nuclear reaction rate in the non-ideal environment we apply both the Monte Carlo technique and approximate analytical calculation of the Feynman diagram using nonrelativistic kinetic Green's functions in the medium which correspond to the generalized energy and momentum distribution functions of interacting particles. We show a possibility to reduce the 12-fold integral corresponding to this diagram to a fivefold integral. This can significantly speed up the computation and control accuracy. Our calculations show that quantum effects significantly influence reaction rates such as p +$^{7} $Be, $^{3} $He +$^{4} $He, p +$^{7} $Li, and $^{12} $C +$^{12} $C. The new reaction rates may be much higher than the classical ones for the interior of the Sun and supernova stars. The possibility to observe the theoretical predictions under laboratory conditions is discussed.
\end{abstract}


\maketitle

\section{\label{intro}Introduction}
The rates of non-resonant nuclear reactions including fusion processes are determined by kinetic energies of the interacting particles in their center-of-mass system as well as by their distribution in energy and momentum. For moderate temperatures the main contribution to the fusion process is expected from particles with energies several times larger than the plasma temperature.

However, it is well known that in dense environments the quantum uncertainty in the energy of particles associated with their frequent collisions leads to disruption of the unambiguous relationship between the energy and momentum of particles \cite{Galitsky1958, Kadanoff:book, Starostin2005UFN}. This results in the appearance of power distributions in the momentum distribution function of particles in dense media. It is particularly interesting to study how these effects in a nonideal plasma contribute to the rates of fusion reactions at moderate plasma temperatures of a few electron volts and densities of about one gram per cubic centimeter

The influence of quantum effects on the equilibrium momentum distribution was investigated by Wigner and others \cite{Wigner1932, Uhlenbeck1932, Landau:book_Stat}, who found the amendment to the Maxwellian distribution, proportional to the square of the Planck constant. Their result was related to the noncommutativity of the operators of kinetic and potential energy of interaction and suggested an effective ``temperature'' increasing with the density of particles.

The first results on the power dependence of the equilibrium distribution function (as $p^{-4} $) for real Bose-particles were obtained by Bogolyubov in his famous work on superfluid weakly nonideal Bose-gas \cite{Bogoliubov1947} (see also \cite{Landau:book_Stat2}). Similarly, one can obtain the momentum distribution for electrons in the superconducting phase at momenta larger than the Fermi momentum. Later, the result of the power law distribution function for momenta larger than the Fermi value was obtained by Belyakov \cite{Belyakov1961} for electrons interacting with short-range impurities. In 1960 Vosko et al \cite{Daniel1960} found the distribution function of electrons above the Fermi momentum, taking into account the exchange interaction of electrons. That function decreased beyond the Fermi jump as $p^{-8} $.

In 1966 Galitsky and Yakimets \cite{Galitsky1966} showed that the equilibrium momentum distribution of particles acquires a power correction term to the Maxwellian function due to quantum effects. This correction is valid at large momenta that exceed the thermal or the Fermi momentum. For the Coulomb interaction potential, the inverse power dependence on the momentum was equal to eight.  In 1975 Kimball proved that in a Coulomb system the asymptotic momentum distribution also includes the eighth degree, regardless of the plasma temperature \cite{Kimball1975}. In the collision of particles of one species in theory there is an additional factor proportional to the correlation function at zero distance between them. This factor may be of the order of 1/2 for electrons colliding with each other due to the Pauli exclusion principle. For heavy particles such as hydrogen or deuterium nuclei this factor is small in the strongly coupled plasma.

In \cite{Starostin2000, Coraddu2004, Starostin2005} the authors proposed a simple model using the Lorentz gas concept in which a light particle is scattered by a heavy impurity particle. It was demonstrated numerically that there must be a significant deviation of the distribution function from the Maxwellian form as the result of the quantum corrections. This fact leads to the power law damping of the distribution function, the exponent is also equal to eight for the screened Coulomb interaction. It has been suggested that this effect should influence the reaction rate for these particles. In particular, it should give the nonexponential temperature dependencies of reaction rate constants for inelastic processes.

In the literature \cite{Zubarev2008, Bahcall2002} there was some criticism, connected with the use of the asymptotical presentation of a single-particle distribution function for calculation of reaction rates, including fusion rates. The problem is that in reality we must use the product of distribution functions over momentum in the laboratory frame for both reacting particles and due to power low tails, not Maxwellian ones, it is hard to perform analytical integration of the reaction cross-section, depending on particles relative momentum in their center of mass.

In this study we simulated the thermonuclear fusion reactions taking into account the impact of this mechanism on the distribution function under conditions that might be realized in a contemporary experiment. The rates of some reactions were calculated. It was shown that in general the reaction rate is determined by the diagram shown in Fig. 1 for the single-particle  nonrelativistic kinetic Green's function or generalized distribution of energy and momentum, which corresponds to the escape process of species $a$ and is reduced, in general, to a ten-fold integral. In other words, in the original expression, one should not perform a simple averaging of the reaction cross-section, depending mainly on the relative momentum of particles (rather than energy) for quantum single-particle momentum distribution function. It is also shown that under certain conditions this integral is reduced to a five-fold integral and in the model case to a three-fold integral. In some cases the last integral can be calculated explicitly and contains contributions from the power type momentum distributions for each of the reacting particles in addition to the classical Maxwellian terms. Each stage of reduction and the corresponding simplification was verified numerically without the use of any simplifications. Some attempts have been made to calculate several fusion reactions appropriate for conditions of the solar plasma in the deep interior. A reasonable accuracy of the simplified explicit estimates for the rate constants was shown for these conditions.
\begin{figure}
\includegraphics[scale=0.65]{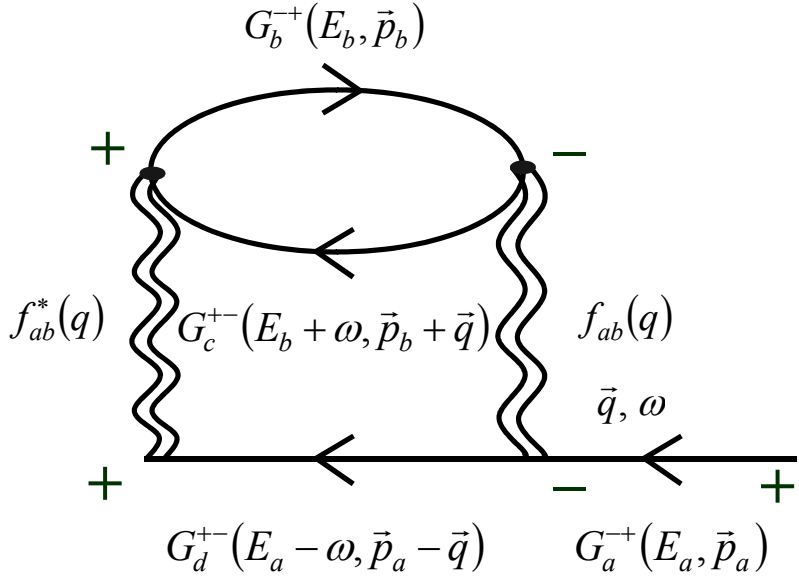}
\caption{\label{fig:1}Reaction rate diagram for $a+b\rightarrow c+d$}
\end{figure}

From our calculations we predict in important circumstances a significant increase in the rate of neutron yield.  In particular, an increase in the rate of d + d reactions may in fact be observable in laboratory testing. We also note possible changes in the reaction rate constant in the plasma of astrophysical objects.

For some reactions the influence of the power corrections was shown to be very significant at relatively low temperatures and high densities of weakly nonideal plasma.

\section{\label{sec:2}Calculation of reaction rates in nonideal plasmas}
The state and properties of a system are found from the generalized distribution function $F\left(E,\vec{p}\right)$ where $E$ and $\vec{p}$ are the energy and momentum of the particles. This function should be defined in a factorized form. Further we use kinetic energy to substitute for the momentum so the generalized function can be written as
\begin{equation}
\label{eq:1}
F\left(E,\varepsilon\right)=n\left(E\right){\kern 1pt}\delta\gamma\left(E-\varepsilon\right) \end{equation}
where $n(E)$ are the occupation numbers and $\delta \gamma (E-\varepsilon )$ is the spectral function, describing the dependence of the generalized distribution function on energy and momentum. Equation (\ref{eq:1}) is the most general representation of the nonrelativistic  kinetic Green's function \cite{Kadanoff:book, Landau:book_Stat2} without any assumptions, in which the width of the Lorentzian spectral function $\delta \gamma$ is the imaginary part of retarded mass operator of the particle in a medium and energy shift is the real part of retarded mass operator. Consequently, the reaction rate follows from the integration which is formally a twelve-fold construction written as (see Fig.~\ref{fig:1})
\begin{eqnarray}
\nonumber
S&=&\frac{1}{4\pi \mu_{ab}^{2} h^{6} } \int_{0}^{\infty }dE_{a} \int d\vec{p}_{a}  \int_{0}^{\infty }dE_{b} \int d\vec{p}_{b}  \int d\omega   \int d\vec{q}\\
\nonumber
&\times&\delta \gamma_{a} \left(E_{a} -\varepsilon_{a} ,\varepsilon_{a} \right)n\left(E_{a} \right)\, \left(1\pm n\left(E_{a} +Q_{a} -\omega \right)\right)\\
\label{eq:2}
&\times& \delta \gamma_{b} \left(E_{b} -\varepsilon_{b} ,\varepsilon_{b} \right)n\left(E_{b} \right)\, \left(1\pm n\left(E_{b} +\omega +Q_{b} \right)\right)\\
\nonumber
&\times& \delta \gamma'_{a} \left(E_{a} +Q_{a} -\omega -\varepsilon_{p_{a} -q} ,\varepsilon_{p_{a} q} \right)\,\\
\nonumber
&\times& \delta \gamma'_{b} \left(E_{b} +\omega +Q_{b} -\varepsilon_{p_{b} +q} ,\varepsilon_{p_{b} +q} \right)\, \left|f\right|^{2}.
\end{eqnarray}
Here the subscript indices $a$ and $b$ designate the reacting species. $E_{a} $ and $p_{a} $ are, respectively, the energy and the momentum of particles $a$, $\varepsilon _{a} $ describes the kinetic energy, $\mu _{ab} $ is the reduced mass, $\omega $ is the interaction energy, $h$ is Planck's constant, and $Q_{a} $ is the energy released in the fusion reaction. The dependencies of the occupation numbers and ``$\pm $'' are subject to proper statistics. Here ``$-$'' refers to fermions and ``+'' to bosons. The actual irreducible dimension of this integral is equal to 10.

For the nonideal plasma, the distribution function contains the Lorentzian which determines the spectral dependence on the kinetic energy:
\begin{equation}
\label{eq:3}
\delta \gamma \left(E-\varepsilon ,\varepsilon \right)=\frac{\gamma \left(E,\varepsilon \right)/\pi }{\left(E-\varepsilon -\Delta \left(E,\varepsilon \right)\right)^{2} +\gamma ^{2} }
\end{equation}
The scattering linewidth in the Lotentz gas model $\gamma (E,\varepsilon )$ is found from
\begin{equation}
\label{eq:4}
\gamma \left(E,\varepsilon \right)=\hbar N\sigma V
\end{equation}
where $N$ is the number density of the scatterers, $\sigma$ is the scattering cross section, and $V$ is the collision velocity, determined by energy $E$.

In Ref. \cite{Starostin1998JETP} it was noted that, for the gaseous medium approximation, when only binary collisions can be considered, the reaction amplitude is just a function of the momenta before and after their reaction. In our further transformations we use the value of the reaction cross section which depends on the energy in the center-of-mass system. The ratio between the amplitude and the cross section has the form
\begin{equation}
\label{eq:5}
\left|f\right|^{2} =\frac{\left|\vec{p}\right|}{\left|\vec{p}-\vec{q}\right|} \sigma _{f} (\varepsilon _{p} )
\end{equation}
where $p$ is the momentum of the reacting species $a$ and $b$ in the center-of-mass system:
\begin{equation}
\label{eq:6}
\vec{p}=\frac{m_b\vec{p}_a-m_a\vec{p}_b}{m_a+m_b} ;\quad \; \quad \varepsilon _{p} =\frac{\left|\vec{p}\right|^{2} }{2\mu _{ab} }
\end{equation}
The reaction cross section is the function of kinetic energy $\varepsilon_{p} $ in the center-of-mass system and may be written as in \cite{Ichimaru1993}, i.e.
\begin{equation}
\label{eq:7}
\sigma_{f}(\varepsilon_p)=\frac{S(\varepsilon _{p} )}{\varepsilon _{p} } \exp \left\{-2\pi \eta {\kern 1pt} (\varepsilon _{p} )\right\}
\end{equation}
where $\eta {\kern 1pt} =Z_{1} Z_{2} e^{2} /\hbar \nu $ is the Sommerfeld factor. It is conventional to use a different form of the cross-section for a non-resonant fusion reaction:
\begin{equation}
\label{eq:8}
\sigma _{f} (\varepsilon _{p} )=\frac{S(\varepsilon _{p} )}{\varepsilon _{p} } \exp \left\{-\pi {\kern 1pt} \sqrt{\frac{E_{G} }{\varepsilon _{p} } } \right\}
\end{equation}
where the Gamow parameter is found from
\begin{equation}
\label{eq:9}
E_{G} =\frac{2\mu _{ab} m_{p} Z_{1}^{2} Z_{2}^{2} e^{4} }{\hbar ^{2} } =4\mu _{ab} \frac{m_{p} }{m_{e} } Z_{1}^{2} Z_{2}^{2} Ry
\end{equation}
with $Ry=m_{e} e^{4} /2\hbar ^{2} $. With a very good accuracy we can approximate it as $Ry=100\mu _{ab} Z_{1}^{2} Z_{2}^{2} $ [keV]. At the same time, the factor $S(\varepsilon_p)$ is weakly dependent on energy $\varepsilon_p$.

In a dense medium with account for the effects of degeneracy the scattering amplitude for nuclear fusion may depend on the total energy \cite{Abrikosov:book}. This leads to corrections proportional to plasma concentration multiplied by the cube of the elastic scattering amplitude, which are low within the gas approximation.

The reaction rate found from (\ref{eq:2}) is a very general definition formulated for this model. The modeling procedure is reduced to calculation of the tenfold integral which is a very complicated task.  Under conditions when plasma becomes ideal, for example for smaller densities, we may write $\gamma \left(E,\varepsilon \right)\to 0$ and $\delta\gamma\left(E-\varepsilon ,\varepsilon \right)$ can be reduced to the delta function.

For reactions with energy release ($Q_{a} >0$), in case of nondegenerate plasma the population numbers are small and $n(E)$ can be neglected as compared to unity. With this simplification the reaction rate obtained from the general equation is the following
\begin{eqnarray}
\label{eq:10}
\nonumber
S&=&\frac{1}{4\pi \mu_{ab}^{2} h^{6}}
\int_{0}^{\infty }dE_{a} \int d\vec{p}_{a}\delta\gamma_{a}\left(E_{a}-\varepsilon_{a} ,\varepsilon_{a} \right)\, n\left(E_{a} \right)\\
%\nonumber\\
&\times&\int_{0}^{\infty }dE_{b} \int d\vec{p}_{b} \delta \gamma_{b} \left(E_{b} -\varepsilon_{b} ,\varepsilon_{b} \right)n\left(E_{b}\right)\\
%\nonumber\\
\nonumber
&\times&\int d\omega  \int d\vec{q} \, \delta \gamma '_{a} \left(E_{a} +Q_{a} -\omega -\varepsilon_{p_{a} -q} ,\varepsilon_{p_{a} q} \right)\, \delta \gamma '_{b} \left(E_{b} +\omega +Q_{b} -\varepsilon_{p_{b} +q} ,\varepsilon_{p_{b} +q} \right)\, \left|f\right|^{2}.
\end{eqnarray}
In this approximation we also neglect the suppression of transmission for Fermi particles or amplification (condensation) for Bose particles. This is a valid assumption because of small populations in both cases.  We further assume that the linewidths of the energy and the kinetic energy distribution profiles, determined by $\delta \gamma '_{a} $ and $\delta \gamma '_{b} $, are small enough to approximate the contours with the Dirac delta functions. With this assumption in equation we come to
\begin{eqnarray}
\nonumber
S&=&\frac{1}{4\pi \mu_{ab}^{2} \left(2\pi \hbar \right)^{6} }
\int_{0}^{\infty }dE_{a} \int d\vec{p}_{a}  n\left(\varepsilon_{a} \right)\delta \gamma_{a} \left(E_{a}-\varepsilon_{a},\varepsilon_{a}\right)\\
\label{eq:11}
&\times& \int_{0}^{\infty }dE_{b} \int d\vec{p}_{b}  n\left(\varepsilon_{b} \right)\delta \gamma_{b} \left(E_{b} -\varepsilon_{b} ,\varepsilon_{b} \right)\left|\vec{p}\right|\sigma \left(\varepsilon_{p} \right)\\
\nonumber
&\times&  \int d\omega  \int d\vec{q} \, \delta \left(E_{a} +Q_{a} -\omega -\varepsilon_{p_{a} -q} \right)\, \delta \left(E_{b} +\omega +Q_{b} -\varepsilon_{p_{b} +q} \right)\, \frac{1}{\left|\vec{p}-\vec{q}\right|}.
\end{eqnarray}
The inner integration over $\omega $ in this equation can be performed as following:
\begin{eqnarray}
\nonumber
I_{q}&=&\int d\vec{q}\int d\omega  \, \delta \left(E_{a} +Q_{a} -\omega -\varepsilon_{p_{a} -q} \right)\, \delta \left(E_{b} +Q_{b} +\omega -\varepsilon_{p_{b} +q} \right) \frac{1}{\left|\vec{p}-\vec{q}\right|}\\
\label{eq:12}
&=&\int d\vec{q}\, \delta \left(E_{a} +Q_{a} +E_{b} +Q_{b} -\varepsilon_{p_{a} -q} -\varepsilon_{p_{b} +q} \right)\frac{1}{\left|\vec{p}-\vec{q}\right|}  .
\end{eqnarray}
Now it is easy to obtain the ratio for the delta function arguments:
\begin{eqnarray}
\label{eq:13}
&&\varepsilon_{\vec{p}_{a}-\vec{q}} +\varepsilon_{\vec{p}_{b} +\vec{q}} =\frac{\left|\vec{p}_{a} -\vec{q}\right|^{2} }{2m_{a} } +\frac{\left|\vec{p}_{b} +\vec{q}\right|^{2} }{2m_{b} }\\
\nonumber
&&=\varepsilon_{a} +\varepsilon_{b} +\frac{q^{2} }{2} \left(\frac{1}{m_{a} } +\frac{1}{m_{b} } \right)-2q\, \left(\frac{p_{a} }{2m_{a} } \cos \left(\vec{p}_{a} \vec{q}\right)-\frac{p_{b} }{2m_{b} } \cos \left(\vec{p}_{b} \vec{q}\right)\right).
\end{eqnarray}
Let us reduce it to the perfect square by introducing the energy in the center-of-mass system:
\begin{equation}
\label{eq:14}
\varepsilon _{\vec{p}_{a} -\vec{q}} +\varepsilon _{\vec{p}_{b} +\vec{q}} =\varepsilon _{a} +\varepsilon _{b} -\frac{p^{2} }{2\mu _{ab} } +\frac{\left(\vec{q}-\vec{p}\right)^{2} }{2\mu _{ab} } =\varepsilon _{a} +\varepsilon _{b} -\varepsilon _{p} +\frac{\left(\vec{q}-\vec{p}\right)^{2} }{2\mu _{ab} }
\end{equation}
so that the integral $I_q$ now takes the form
\begin{eqnarray}
\label{eq:15}
I_{q}&=&\int d\vec{q} \, \delta \left(E_{a} +Q_{a} +E_{b} +Q_{b} -\varepsilon_{p_{a} -q} -\varepsilon_{p_{b} +q} \right) \frac{1}{\left|\vec{q}-\vec{p}\right|}\\
\nonumber
&=&\int_{}^{}d\vec{q} \, \delta \left(E_{a} +E_{b} +Q_{a} +Q_{b} -\varepsilon_{p_{a} } -\varepsilon_{p_{b} } +\varepsilon_{p} -\frac{\left(\vec{q}-\vec{p}\right)^{2} }{2\mu_{ab} } \right)\frac{1}{\left|\vec{q}-\vec{p}\right|}
\end{eqnarray}
Let us note that variable $\varepsilon_q$ as well as momentum $\vec{p}$ is determined by the momenta of colliding particles and does not depend on $\vec{q}$. Changing to the variable $\vec{s}=\vec{q}-\vec{p}$ we get
\begin{eqnarray}
\nonumber
I_{q}&=&\int_{}^{}d\vec{s} \; \delta \left(E_{a} +E_{b} +Q_{a} +Q_{b} -\varepsilon_{p_{a} } -\varepsilon_{p_{b} } +\varepsilon_{p} -\frac{\left(\vec{s}\right)^{2} }{2\mu_{ab} } \right) \frac{1}{\left|\vec{q}-\vec{p}\right|}\\
\label{eq:16}
&=&4\pi \int_{0}^{\infty }s^{2} ds \; \delta \left(E_{a} +E_{b} +Q_{a} +Q_{b} -\varepsilon_{p_{a} } -\varepsilon_{p_{b} } +\varepsilon_{p} -\frac{\left(\vec{s}\right)^{2} }{2\mu_{ab} } \right) \frac{1}{\left|\vec{s}\right|}\\
\nonumber
&=&4\pi \mu_{ab} \int_{0}^{\infty }d\varepsilon_{s}  \delta \left(E_{a} +E_{b} +Q_{a} +Q_{b} -\varepsilon_{p_{a} } -\varepsilon_{p_{b} } +\varepsilon_{p} -\varepsilon_{s} \right)=4\pi \mu_{ab}.
\end{eqnarray}
The integration in the latter equation was carried out with the use of the variable $\varepsilon _{s} =\; {\kern 1pt} |\vec{s}|^{2} /2\mu _{ab} $.

Now we substitute the result into equation (\ref{eq:11}). For the relative velocity of the colliding particles we need to make an account for $V_{ab} =V_{a} -V_{b} ={\kern 1pt} |\vec{p}|/\mu _{ab} $. As the result we come to
\begin{eqnarray}
\nonumber
S&=&\frac{1}{4\pi \mu_{ab}^{2} } \int_{0}^{\infty }\frac{dE_{a} }{\left(2\pi\hbar\right)^{3}}\int d\vec{p}_{a}\,n\left(E_{a}\right)\delta_\gamma\left(E_{a}-\varepsilon_{a} \right)\int_{0}^{\infty}\frac{dE_{b}}{\left(2\pi\hbar\right)^{3}
}\\
\label{eq:17}
&\times& \int d\vec{p}_{b}\,n\left(E_{b} \right)\delta_\gamma \left(E_{b} -\varepsilon_{b} \right)\, \mu_{ab} \frac{\left|\vec{p}\right|}{\mu_{ab} }\,\sigma \left(\varepsilon_{p} \right)\,4\pi \mu_{ab}\,\\
\nonumber
&=&\int_{0}^{\infty }\frac{dE_{a} n\left(E_{a} \right)}{\left(2\pi \hbar \right)^{3} } \int d\vec{p}_{a}  \, \delta_\gamma \left(E_{a} -\varepsilon_{a} \right)\int_{0}^{\infty }\frac{dE_{b} n\left(E_{b} \right)}{\left(2\pi \hbar \right)^{3} } \int d\vec{p}_{b}  \, \delta_\gamma \left(E_{b} -\varepsilon_{b} \right)V_{ab} \left(\varepsilon_{p} \right)\,\sigma_{f}\left(\varepsilon_{p} \right).
\end{eqnarray}
If we use the apparent ratio for the number densities $N_{a} $ and $N_{b} $ this equation can be transformed to the standard form, i.e.
\begin{equation}
\label{eq:18}
S=\left\langle V_{ab} \sigma \right\rangle N_{a} N_{b}
\end{equation}
Numerical simulation of a fusion reaction in the light of the above models and the calculation of the reaction rates with (\ref{eq:2}) and (\ref{eq:17}) as well as comparison of the results confirmed the correctness of the approximations used in our approach. Table \ref{T:1} shows the constants for the model reaction $\langle \sigma V_{ab} \rangle $ for values of particles concentration $N$ and temperature $T=2$ eV in the region where quantum effects are important. The reaction constants $\langle \sigma V_{ab} \rangle $ obtained with the numerical model in which we used equation (2) is designated as $\langle V_{ab} \sigma \rangle \_10$ and the results obtained with equation (\ref{eq:17}) are denoted as $\langle V_{ab} \sigma \rangle \_5$. The detailed simulation algorithm in the Monte Carlo method is described in the appendix. These results were compared with the analytical estimations based on (73) which is obtained later in this paper. Calculations were performed for the astrophysical factor $S\left(0\right)=241$ keV barns and the Gamow energy was determined in accordance with the equation (\ref{eq:9}) as $E_{G} =100$ keV.
\begin{table}
  \centering
    \begin{tabular}{|c|c|c|c|}
    \hline
    & $\,\bf N=10^{21}\,cm^{-3}\,$ &$\,\bf N=10^{22}\,cm^{-3}\,$ & $\,\bf N=10^{23}\,cm^{-3}\,$ \\
    \hline
    $\textrm{Analytic}$ & $1.5\times10^{-29}$ & $1.5\times10^{-28}$   &  $1.5\times10^{-27}$ \\
    \hline
    $\left\langle V_{ab} \sigma \right\rangle\_5$ & $1.3\times10^{-29}$ & $2.0\times10^{-28}$ & $3.2\times10^{-27}$ \\
    \hline
    $\left\langle V_{ab} \sigma \right\rangle\_10$ & $0.99\times10^{-29}$ & $2.1\times10^{-28}$ & $2.7\times 10^{-27}$ \\
    \hline
    \end{tabular}
\caption{\label{T:1}}
\end{table}

\section{\label{sec:3}Kinetic energy distribution function with quantum corrections}
For the nondegenerate plasma the occupation numbers should have Maxwellian distribution over energy at high temperatures:
\begin{equation}
\label{eq:19}
n(\varepsilon )=\frac{2}{\sqrt{\pi } \left(kT\right)^{3/2} } \exp \left\{-\frac{\varepsilon }{kT} \right\}
\end{equation}
At the same time the energy distribution function could be derived by integrating the generalized distribution function as shown here:
\begin{eqnarray}
\label{eq:20}
f(\varepsilon )&=&\int_{0}^{\infty }f\left(E,\varepsilon \right)\, dE =\int_{0}^{\infty }n\left(E\right)\, \delta_{\gamma } (E-\varepsilon )\, dE\\
\nonumber
&=&\frac{1}{\pi } \int_{0}^{\infty }n\left(E\right)\frac{\gamma \left(E,\varepsilon \right)}{\left(E-\varepsilon -\Delta \left(E,\varepsilon \right)\right)^{2} +\gamma \left(E,\varepsilon \right)^{2} }\,dE.
\end{eqnarray}
For the values of the kinetic energy in the range of $\varepsilon \le E_{0} $, where the threshold energy is significantly greater than the linewidth of the Lorentz function, i.e. $\gamma <<E_{0} $, the Lorentzian can be represented by $\delta $--function with a high accuracy. Therefore, for this energy range the kinetic energy distribution function is $f(\varepsilon )=n(\varepsilon )$. If the kinetic energy is $\varepsilon >E_{0} $, then in order to calculate the integral one should split the integration interval in Eq. (\ref{eq:20}) into two regions, i.e. the region of small energies of the order of the plasma temperature $E_{0}~{\rm :}~3kT...5kT$ and the rest of the interval. In the region of $E>E_{0} $ the Lorentzian can be approximated by the delta function. In the region of small $E$ with account for $\gamma <<kT$ in the asymptotic limit of $\varepsilon >>kT$ the denominator of the Lorentzian is approximately $\varepsilon ^{2}$:
\begin{eqnarray}
\nonumber
f\left(\varepsilon \right)&=&\int_{0}^{E_{0} }n\left(E\right)\frac{{\gamma \left(E,\varepsilon \right) \mathord{\left/{\vphantom{\gamma \left(E,\varepsilon \right) \pi }}\right.\kern-\nulldelimiterspace} \pi } }{\left(E-\varepsilon -\Delta \left(E,\varepsilon \right)\right)^{2} +\gamma \left(E,\varepsilon \right)^{2} }\,dE\\
\nonumber
\\
%\nonumber
\label{eq:21}
&+&\int_{E_{0} }^{\infty }n\left(E\right)\frac{{\gamma \left(E,\varepsilon \right) \mathord{\left/{\vphantom{\gamma \left(E,\varepsilon \right) \pi }}\right.\kern-\nulldelimiterspace} \pi } }{\left(E-\varepsilon -\Delta \left(E,\varepsilon \right)\right)^{2} +\gamma \left(E,\varepsilon \right)^{2} }\,dE\\
\label{eq19}
\nonumber\\
\nonumber
&=&\frac{2}{\sqrt{\pi } \, \left(kT\right)^{{3 \mathord{\left/{\vphantom{3 2}}\right.\kern-\nulldelimiterspace} 2} } } \left[\int_{0}^{E_{0} }\exp \left\{-\frac{E}{kT} \right\}\frac{\gamma \left(E,\varepsilon \right)/\pi }{\varepsilon ^{2} }\,dE +\int_{E_{0} }^{\infty }\exp \left\{-\frac{E}{kT} \right\}\delta \left(E-\varepsilon \right)dE \right]\\
\nonumber
&&\\
\nonumber
&=&f_{qt} \left(\varepsilon \right)+f_{0} \left(\varepsilon \right).
\end{eqnarray}
The classic expression for the distribution function is represented by the term
\begin{equation}
\label{eq:22}
f_{0} \left(\varepsilon \right)=\frac{2}{\sqrt{\pi } \left(kT\right)^{3/2} } \exp \left\{-\frac{\varepsilon }{kT} \right\}
\end{equation}
In the range of small $E$ it is necessary to use the Coulomb cross section:
\begin{equation}
\label{eq:23}
\sigma_t(\varepsilon_p)=\frac{2\pi e^{4} Z_{a}^{2} Z_{l}^{2} }{\varepsilon _{p}^2}
\end{equation}
and the expression for width of the Lorentz function which is conditioned by scattering of particle $a$ on plasma particles $l$:
\begin{equation}
\label{eq:24}
\gamma _{al} =\hbar N_{l} \sigma _{t} (\varepsilon _{p} )V_{al}
\end{equation}
In the center-of-mass system we write
\begin{equation}
\label{eq:25}
\varepsilon _{p} =\mu _{al} \left(\frac{\varepsilon _{a} }{m_{a} } +\frac{\varepsilon _{l} }{m_{l} } -2\sqrt{\frac{\varepsilon _{a} }{m_{a} } \frac{\varepsilon _{l} }{m_{l} } } \cos \left(\vec{p}_{a} ,\vec{p}_{l} \right)\right)
\end{equation}
If we consider the Lorentz gas approximation, i.e. $m_a\ll m_l$, then
\begin{equation}
\label{eq:26}
\varepsilon _{a} /m_{a} >>\varepsilon _{l} /m_{l} ,\frac{\varepsilon _{p} }{\mu _{al} } \approx \frac{\varepsilon _{a} }{m_{a} }
\end{equation}
where
\begin{equation}
\label{GrindEQ__27_}
\varepsilon _{a} =\frac{\left|\vec{p}_{a} \right|^{2} }{2m_{a} }
\end{equation}
\begin{equation}
\label{GrindEQ__28_}
V_{ab} =\sqrt{\frac{2\varepsilon _{p} }{\mu _{ab} } } \approx \sqrt{\frac{2\varepsilon _{a} }{m_{a} } } \quad or\quad V_{ab} \approx \sqrt{\frac{2E_{a} }{m_{a} } }
\end{equation}
The cross section for Coulomb scattering of particles $a$ by particles $l$:
\begin{equation}
\label{GrindEQ__29_}
\sigma _{al} \left(\varepsilon \right)=\frac{2\pi e^{4} Z_{a}^{2} Z_{l}^{2} }{\varepsilon _{p{\kern 1pt} al}^{2} } =\frac{2\pi e^{4} Z_{a}^{2} Z_{l}^{2} m_{a}^{2} }{\mu _{al}^{2} \varepsilon _{a}^{2} }
\end{equation}
\begin{equation}
\label{GrindEQ__30_}
\gamma _{al} =\hbar N_{l} \frac{2\pi e^{4} Z_{a}^{2} Z_{l}^{2} }{\varepsilon _{a}^{2} } \frac{m_{a}^{2} }{\mu _{al}^{2} } \sqrt{\frac{2E_{a} }{m_{a} } }
\end{equation}
where $\frac{m_{a} }{\mu _{al} } =\frac{m_{a} }{m_{a} m_{l} } \left(m_{a} +m_{l} \right)\approx 1$        for $m_{a} <<m_{l} $. % (31)

In a multicomponent medium the line width is determined by the sum of the contributions from different sorts of scatterers (strictly speaking, the sum over $l$ must take into account the particles different from species $a$, according to the Lorentz model):
\begin{equation}
\label{GrindEQ__32_}
\gamma _{a} =\sum _{l} \gamma _{al} =\frac{2\pi \hbar e^{4} Z_{a}^{{\kern 1pt} 2} }{\varepsilon _{a}^{2} } \sqrt{\frac{2E_{a} }{m_{a} } } \sum _{l} N_{l} Z_{l}^{{\kern 1pt} 2} \frac{m_{a}^{2} }{\mu _{al}^{2} } =\frac{2\pi \hbar e^{4} Z_{a}^{{\kern 1pt} 2} \Sigma _{al} }{\varepsilon _{a}^{2} } \sqrt{\frac{2E_{a} }{m_{a} } }
\end{equation}
Here we introduced the following notation:
\begin{equation} \label{GrindEQ__33_} \Sigma _{al} =\sum _{l} N_{l} Z_{l}^{{\kern 1pt} 2} \frac{m_{a}^{2} }{\mu _{al}^{2} }  \end{equation}
\begin{figure}
\includegraphics[scale=0.75]{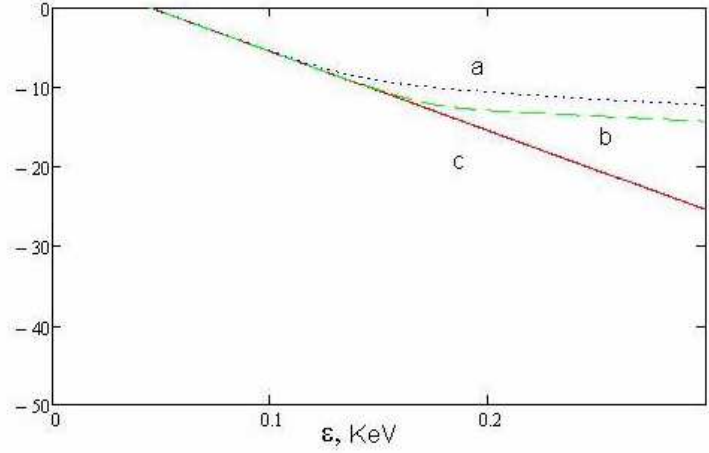}
\caption{\label{fig:2}The distribution function of the kinetic energy in the Lorentz gas for temperature 0.01 KeV (a), obtained from numerical simulation of the generalized distribution function (b) and the Maxwellian distribution (c). }
\end{figure}
\begin{figure}
\includegraphics[scale=0.83]{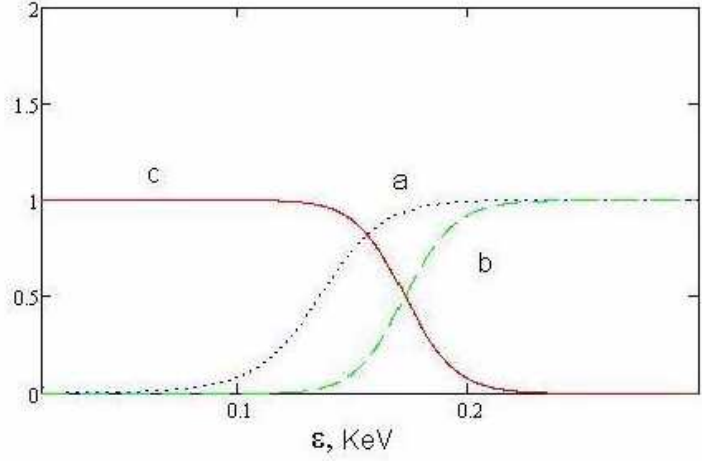}
\caption{\label{fig:3}The share of the quantum correction to the full distribution function in the approximation of the Lorentz gas for temperature 0.01 KeV (a), the same quantity for the generalized distribution function, obtained from numerical simulation (b), the share of the Maxwellian part to the full distribution function (c). }
\end{figure}
\begin{figure}
\includegraphics[scale=0.75]{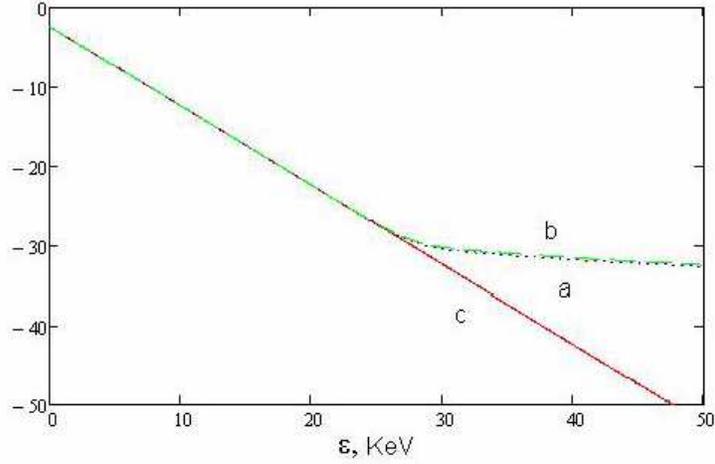}
\caption{\label{fig:4}The same as in Figure \ref{fig:2} for temperature 1.0 KeV.}
\end{figure}
\begin{figure}
\includegraphics[scale=0.75]{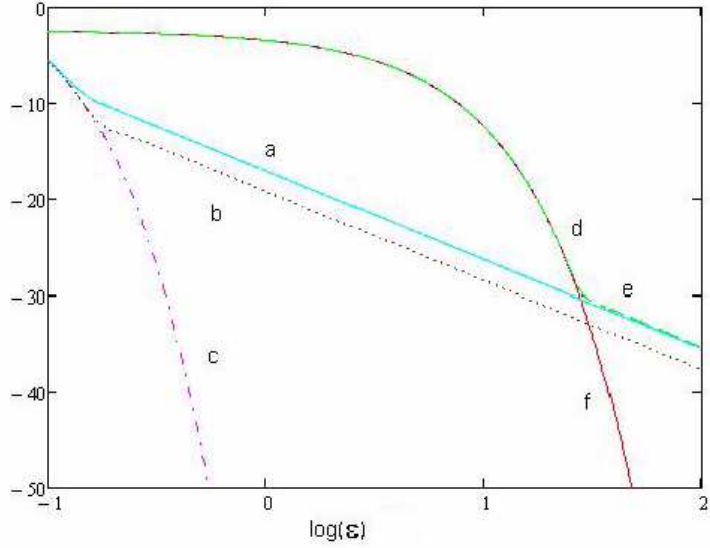}
\caption{\label{fig:5}Curves (a)-(c) the same as in Figure \ref{fig:2} (temperature 0.01 KeV), curves (d)-(f) respectively the same functions of temperature 1.0 KeV.}
\end{figure}

In order to calculate the ``tail'' of the distribution function, we need to substitute the line width from Eq. (\ref{GrindEQ__30_}) into integral (\ref{eq:21}) and write
\begin{eqnarray}
\nonumber
f_{qt} \left(\varepsilon _{a} \right)&=&\frac{2}{\sqrt{\pi } \, \left(kT\right)^{3/2}}
\int_{0}^{\infty }\exp \left\{-\frac{E_{a} }{kT} \right\}\frac{\gamma \left(E_{a} ,\varepsilon _{a} \right)}{\pi \varepsilon _{a}^{2} } dE_{a}\\
\nonumber
\\
\label{eq:34}
&=&\frac{2}{\sqrt{\pi } \, \left(kT\right)^{3/2}}
\int_{0}^{\infty }\exp \left\{-\frac{E_a}{kT} \right\}
\frac{2\sqrt{2}\pi \hbar e^{4} Z_{a}^{2} }{\varepsilon _{a}^{2}\sqrt{m_a}}
\frac{\sqrt{E_a} \Sigma _{al}}{\pi \varepsilon_a^2} dE_{a}\\
\nonumber
\\
\nonumber
&=&\frac{2}{\sqrt{\pi } \, \left(kT\right)^{3/2}} \frac{\sqrt{2\pi } \hbar e^{4} Z_{a}^{2}\Sigma _{al} }{\sqrt{m_{a} } } \frac{1}{\varepsilon _{a}^{4} } \left(kT\right)^{3/2} =\frac{2\sqrt{2} \hbar e^4 Z_a^2}{\sqrt{m_a}}\frac{\Sigma_{al} }{\varepsilon_a^4}.
\end{eqnarray}
The influence of the correction to the distribution function is significant only in the asymptotic region of its argument which is the kinetic energy. It should be noted that at low values of the kinetic energy $\varepsilon \le E_{0} $ the distribution function is determined by the classical expression, i.e., the Maxwellian function (\ref{eq:22}). For the temperature of 10 eV and the concentration of interacting particles (deuterium ions) $N=10^{23} $ cm$^{-3} $ the value of the energy threshold parameter can be defined as $E_{0} =5kT$. In this case, taking into account equation (\ref{GrindEQ__30_}) one can male the following estimation:
\begin{equation}
\label{GrindEQ__35_}
\gamma /E_{0} =\frac{\hbar N\sigma V}{E_{0} } =\frac{2\pi \hbar Ne^{4} }{(kT)^{2} } \sqrt{\frac{2kT}{m} } \frac{1}{E_{0} } \approx 5\times 10^{-3}
\end{equation}
Figure 2 shows the contribution of the quantum-tail correction in the approximation of the Lorentz gas in the full distribution function.  Figure \ref{fig:2} shows the Maxwellian distribution functions with a correction that takes into account the quantum effects in the approximation of the Lorentz gas and obtained in the analytical form (\ref{eq:21}), as well as the distribution function of the kinetic energy, obtained by numerical integration of the generalized distribution function.

In Figure \ref{fig:3} this contribution is presented as the ratio $f_{qt}(E)/f(E)$ in equation (\ref{eq:21}), where - curve (a), the same value for the distribution of kinetic energy, obtained by numerical integration of the generalized distribution function - curve (b), as well as the contribution of the Maxwell distribution in the full distribution function kinetic energy, the ratio of the formula (\ref{eq:21}) - curve (c). The simulation was performed under conditions of a shock compression of deuterium in the titanium matrix with the density of atoms of $5.7\times 10^{22} $ cm$^{-3} $ at a temperature of $0.01$ keV. As one can see the energy is less than $0.1$ keV the total distribution function is determined by the Maxwellian term. At energies higher than $0.2$ keV the contribution of the quantum correction to the total distribution function becomes crucial. Quantum correction obtained in the approximation of the Lorentzian gas as well as obtained by numerical integration of the generalized distribution functions, represented by curves (a) and (b) respectively.

Figure \ref{fig:4} shows the same distribution function as in Fig.~\ref{fig:2}, but at a temperature of 1.0 keV. As one can see from comparison of these figures, with increasing temperature the quantum effect comes into play at higher energy.

Figure \ref{fig:5} shows the energy distribution for different temperatures: 0.01 and 1.0 keV, curves shown are collected from Figs. \ref{fig:2} and \ref{fig:4}. Given that the characteristic energies are of different orders, we had to switch to a logarithmic scale. It is clear that in the asymptotic region the tails of the distribution function do not depend on temperature and are practically identical. From this pictures it may be estimated the percentage of deuterium ions in the asymptotic region of distribution function over momentum. This number is much larger then correspondent number in Maxwellian distribution, which is considered for calculating fusion rate constant in classical approach.

\section{\label{sec:4}Comparison of Kimball's approach and the Lorentz gas model}
The equations for the power-tail distribution function of particles momentum in the Lorentz model can be compared with the rigorous result obtained with the use of the Kimball's approach. Considering the repulsive Coulomb interaction between two particles with charges $Z_{a} $ and $Z_{b} $, and reduced mass $m_{ab} $ from the Schrodinger equation it follows:
\begin{equation}
\label{eq:36}
\left(-\frac{\hbar ^{2} }{2\mu _{ab} } \nabla _{ab}^{2} +\frac{Z_{a} Z_{a} e^{2} }{r_{ab} } \right)\psi =F
\end{equation}
where $F$ does not contain singularities at small $r$. At short distances one can solve the equation (extraction of the singularity at small $r$)
\begin{equation}
\label{eq:37}
\psi =f\left(1+\frac{\left|r\right|}{a_{ab} } \right)
\end{equation}
where $a_{ab} $ is the Bohr radius:
\begin{equation}
\label{eq:38}
a_{ab} =\frac{\hbar ^{2} }{\mu _{ab} Z_{a} Z_{b} e^{2} }
\end{equation}
Using this solution one can come to the distribution function by calculating the Fourier transform of the product of the solutions with singularities at the particle species $a$ approaching particles $l$ and $m$. In the final expression points $l$ and $m$ are to tend to each other. As a result, we obtain the asymptotic momentum distribution function in the following form
\begin{equation}
\label{GrindEQ__39_}
f_{al} \left(p\right)=\frac{64\pi ^{2} \mu _{al}^{2} }{\hbar ^{3} \hbar ^{4} } \frac{Z_{a}^{2} Z_{l}^{2} e^{4} N_{l} }{k^{8} }
\end{equation}
It is easy to see that these asymptotics in the framework of the Lorenz model and the Kimball's approach are in good agreement. This agreement takes place independently of the mass of the particles -- whether electrons or nuclear particles -- and whether or not they are degenerate.

It is interesting to note that. in contrast to the Lorenz model. the Kimball's formulas are valid for an arbitrary mass ratio of the colliding particles. For the scattering of particles of species $a$ on one another we can get, using  Kimball's method, the following:
\begin{equation}
\label{GrindEQ__40_}
f_{aa} \left(p\right)=\frac{16\pi ^{2} \mu _{aa}^{2} }{\hbar ^{7} } \frac{Z_{a}^{4} e^{4} N_{l} }{k^{8} } g_{aa} (0)
\end{equation}
In a nonideal plasma these contributions are small in proportion to the ion-ion correlation function. Contribution to the quantum asymptotic of the distribution function for protons and other ions at expense of their scattering on neutrals and electrons is small due to small cross sections and the square of the reduced mass.

One can compare the asymptotics of the distribution function obtained by Kimball and within the Lorentz model for the interaction potential, which has a singularity near zero distance between the particles:
\begin{equation}
\label{GrindEQ__41_}
U=\frac{C}{r^{n} } \quad \quad \left(n\le 3\right)
\end{equation}
The Schrodinger equation with such a potential is written as
\begin{equation}
\label{GrindEQ__42_}
\left(-\frac{\hbar ^{2} }{2\mu _{ab} } \nabla _{ab}^{2} +\frac{C}{r^{n} } \right)\psi =F \end{equation}
and has a solution
\begin{equation}
\label{GrindEQ__43_}
\psi \left(r\right)=\psi _{0} \left(1-\frac{2\mu _{al} C}{\hbar ^{2} \left(n-2\right)\left(3-n\right)r^{n-2} } \right)
\end{equation}
After calculating the square of the Fourier transform of this solution we obtain
\begin{equation}
\label{GrindEQ__44_}
n_{p} {\rm \sim }\frac{1}{k^{10-2n} }
\end{equation}
At the same time, from the expression for the quantum correction to the Maxwellian distribution we get
\begin{equation}
\label{GrindEQ__45_}
f_{p} {\rm \sim }\frac{\hbar N_{l} T^{3/2} \sigma _{t} (l)\sqrt{\pi } }{2\pi \varepsilon _{p}^{2} \sqrt{2\mu _{al} } } {\rm \sim }\frac{1}{k^{4} } \left|\int  e^{-i\vec{k}\cdot \vec{r}} U\left(r\right)d^{3} r\right|^{2}
\end{equation}
where
\begin{equation}
\label{GrindEQ__46_}
\sigma _{t} {\rm \sim }k^{2n-6}
\end{equation}
Calculating the scattering amplitude on the potential of this type in the Born approximation we obtain:
\begin{equation}
\label{GrindEQ__47_}
f_{n} {\rm \sim }\frac{\sigma _{t} }{\varepsilon _{p}^{2} } {\rm \sim }\frac{1}{k^{10-2n} }  \end{equation}
which is in agreement with (\ref{GrindEQ__44_}).

In addition to the agreement, as noted above, of the asymptotic expression (\ref{GrindEQ__39_}) to the limit obtained by Kimball for the electron momentum distribution function due to their interaction with ions, with the results of the Lorentz model, this treatment now allows us to generalize the result for the distribution function of heavy particles such as deuterons interacting with heavy ions, as soon as they satisfy the Schrodinger equation and the Coulomb law at distances of the order of the Bohr radius (\ref{eq:38}). Note that the result obtained by Kimball does not depend on the ratio of masses of interacting particles, since the interaction is considered in the center-of-mass frame. Thus, we can conclude that the power law for the distribution function in the asymptotic region holds not only where the Lorentz model is applicable to interacting particles, but also for arbitrary particles. For example, it holds also for the interaction of deuteron ions with hydrogen ions. This asymptotic behavior does not depend on the temperature, so the Kimball's theory for electron distribution function and Lorentz gas model  considered here coincides in spite of great differences of physical objects, i.e., metals near zero degree in the first case and dense plasmas of 1-1000 eV temperature in the second.

It may also be noted that if electrons are localized within their Bohr radius, with an uncertainty in their momentum and kinetic energy of the order of 13.6 eV, then for deuterons, similarly localized to a Bohr radius (\ref{eq:38}), the scale of uncertainty of their kinetic energy will be of the order of 50 keV, which is on the scale of energy necessary for fusion to occur.

\section{\label{sec:5}Reaction rate for the model distribution functions}
The reaction rate for the case of the nondegenerate plasma at sufficiently high temperature is determined by averaging the interaction frequency of the particles over their distribution functions. The equation to calculate the reaction frequency is reduced to finding
\begin{equation}
\label{GrindEQ__48_}
\left\langle \sigma V\right\rangle =\int  \int  f\left(\varepsilon _{a} \right)f\left(\varepsilon _{b} \right)V_{ab} \sigma _{f} \left(\varepsilon _{p} \right)d^{3} p_{a} d^{3} p_{b}  \end{equation}
Here $V_{ab} $ is the relative velocity of particles $a$ and $b$, $\varepsilon _{p} $ is the energy of the particles in the center-of-mass system.

The relation between the energy of the particles in the center of mass coordinate system and in the laboratory system has the form:
\begin{equation}
\label{GrindEQ__49_}
\varepsilon _{p} =\mu _{ab} \left(\frac{\varepsilon _{a} }{m_{a} } +\frac{\varepsilon _{b} }{m_{b} } -2\sqrt{\frac{\varepsilon _{a} \varepsilon _{b} }{m_{a} m_{b} } } \cos \left(\vec{p}_{a} ,\vec{p}_{b} \right)\right)
\end{equation}
In the laboratory system we have
\begin{equation}
\label{GrindEQ__50_}
V_{ab} =\left|\vec{V}_{a} -\vec{V}_{b} \right|=\sqrt{\frac{2\varepsilon _{p} }{\mu _{ab} } } =\sqrt{2} \sqrt{\frac{\varepsilon _{a} }{m_{a} } +\frac{\varepsilon _{b} }{m_{b} } -2\sqrt{\frac{\varepsilon _{a} \varepsilon _{b} }{m_{a} m_{b} } } \cos \left(\vec{p}_{a} ,\vec{p}_{b} \right)}
\end{equation}
Using equation (\ref{GrindEQ__48_}) for calculating the reaction frequency might be easier if one reduces the integral's dimension by performing the angular integration:
\begin{equation}
\label{GrindEQ__51_}
\left\langle \sigma V\right\rangle =2(2\pi )^{2} \int _{0}^{\infty } dp_{a} p_{a}^{2} f(\varepsilon _{a} )\int  dp_{b} p_{b}^{2} f\left(\varepsilon _{b} \right)\int _{-1}^{1} d\cos \theta _{ab} V_{ab} \sigma _{f} (\varepsilon _{p} )
\end{equation}
Using the same approximation for the distribution function (\ref{eq:21}), i.e. marking out the power asymptotics, we can rearrange the equation to form convenient for standard numeric integration. The reaction frequency is then split into four terms, which correspond to the terms of the distribution function in the region of small energies and in the asymptotic region.
\begin{eqnarray}
\nonumber
\left\langle\sigma V\right\rangle&=&2(2\pi)^{2}
\int_{0}^{\infty}dp_{a}p_{a}^{\,2} \left(f_{0} \left(\varepsilon_{a} \right)+f_{qt} \left(\varepsilon_{a} \right)\right)
\int_{0}^{\infty}dp_{b} p_{b}^{\,2} \left(f_{0} \left(\varepsilon_{b} \right)+f_{qt} \left(\varepsilon_{b} \right)\right)
\int_{-1}^{1}d\cos\theta_{ab}V_{ab}\sigma_{f}(\varepsilon_p)\\
\nonumber
&=&2(2\pi)^2
\int_{0}^{\infty}dp_{a}p_{a}^{\,2}f_{0}(\varepsilon_a)
\int_{0}^{\infty}dp_{b}p_{b}^{\,2}f_{0}(\varepsilon_b)
\int_{-1}^{1}d\cos\theta_{ab}V_{ab}\,\sigma_{f}(\varepsilon_p)\\
&+&2\left(2\pi \right)^{2}
\int_{0}^{\infty}dp_{a}p_{a}^{\,2}f_{0}(\varepsilon_a)
\int_{0}^{\infty}dp_{b}p_{b}^{\,2}f_{qt}(\varepsilon_b)
\int_{-1}^{1}d\cos\theta_{ab}V_{ab}\,\sigma_{f}(\varepsilon_p)\\
\nonumber
&+&2(2\pi)^2
\int_{0}^{\infty}dp_{a}p_{a}^{\,2}f_{qt}(\varepsilon_a)
\int_{0}^{\infty}dp_{b}p_{b}^{\,2}f_{0}(\varepsilon_b)
\int_{-1}^{1}d\cos\theta_{ab}V_{ab}\,\sigma_f(\varepsilon_p)\\
\nonumber
&+&2(2\pi)^2
\int_{0}^{\infty }dp_{a} p_{a}^{\,2}f_{qt}(\varepsilon_a)
\int_{0}^{\infty }dp_{b} p_{b}^{\,2}f_{qt}(\varepsilon_b)
\int_{-1}^{1}d\cos\theta_{ab}V_{ab}\,\sigma_{f}(\varepsilon_p)
\end{eqnarray}
This equation for the reaction rate constant can be used in calculations of fusion reaction rates. Along with equations (\ref{eq:2}) and (\ref{eq:17}) this expression is the next order approximation suitable for numeric modeling of fusion reactions.

We change the variables of integration from the momenta to energies and substitute expression (\ref{eq:34}) for the asymptotic distribution function. As a result we obtain:
\begin{eqnarray}
\nonumber
\left\langle\sigma V\right\rangle &=&\frac{2}{\pi \left(kT\right)^{3} }
\int_{0}^{\infty }d\varepsilon_{a} \sqrt{\varepsilon_{a} }  \exp \left\{-\frac{\varepsilon _{a} }{kT} \right\}
\int_{0}^{\infty }d\varepsilon_{b} \sqrt{\varepsilon_{b} }\exp\left\{-\frac{\varepsilon _{b} }{kT} \right\}
\int_{-1}^{1}d\cos\theta_{ab} V_{ab}\,\sigma_{f}\left(\varepsilon _{p} \right)\\
\nonumber
&&\\
\nonumber
&+&\frac{2\sqrt{2\pi }\hbar e^{4} Z_{b}^{2} \Sigma _{bl} }{\pi \left(kT\right)^{3/2}\sqrt{m_{b}}}
\int_{0}^{\infty }d\varepsilon_{a}\sqrt{\varepsilon _{a} }\exp \left\{-\frac{\varepsilon _{a} }{kT} \right\}
\int_{0}^{\infty }\frac{d\varepsilon _{b} }{\varepsilon _{b}^{7/2}}
\int_{-1}^{1}d\cos \theta_{ab} V_{ab}\,\sigma _{f}\left(\varepsilon _{p}\right)\\
\nonumber
\\
\nonumber
&+&\frac{2\sqrt{2\pi }\hbar e^{4} Z_{a}^{2} \Sigma_{al} }{\pi \left(kT\right)^{3/2}\sqrt{m_{a} } } \int_{E_0}^{\infty }\frac{d\varepsilon_a}{\varepsilon_a^{7/2}}
\int_{0}^{\infty }d\varepsilon _{b} \sqrt{\varepsilon _{b} }\exp \left\{-\frac{\varepsilon _{b}}{kT} \right\} \int_{-1}^{1}d\cos\theta _{ab}V_{ab}\,\sigma_{f}\left(\varepsilon _{p} \right)\\
\nonumber
\\
\label{eq:53}
&+&\frac{\left(2\hbar e^{4}Z_{a}Z_{b}\right)^{2}\Sigma_{al}\Sigma _{bl}}{\sqrt{m_{a} m_{b} }}
\int_{E_0}^{\infty } \frac{d\varepsilon _{a} }{\varepsilon _{a}^{7/2}}
\int_{E_0}^{\infty }\frac{d\varepsilon_{b} }{\varepsilon _{b}^{7/2}}
\int_{-1}^{1}d\cos \theta _{ab} V_{ab}\,\sigma _{f} \left(\varepsilon _{p} \right)
\end{eqnarray}
Further calculations of constants for the reaction with the use of this model are denoted as $\left\langle V_{ab} \sigma \right\rangle \_ 3$. The last formula takes into account that the part of the distribution function conditioned by the quantum effects is $f_{qt} (\varepsilon )=0$ in the range of low kinetic energies.

Equation (\ref{eq:53}) can be analyzed to estimate the causes and conditions under which the asymptotic region of the energy distribution functions gives a determining contribution to the rate of fusion reactions.  The integrand in the first term of (\ref{eq:53}) is a series of factors that are notably different from zero in different domains of their arguments. The values of the Maxwellian distribution function are $\sim 1$ for the energies not too much higher than the temperature. In the asymptotic region these functions decrease exponentially. The fusion cross section is exponentially small at low energies and reaches its maximum for the energies of the order of hundreds of keV, as, for example, happens in the synthesis of the deuteron. Therefore, at low temperatures the first term in (\ref{eq:53}) becomes small. Note that it is the one which determines the reaction rate in the model currently accepted for computation.

The second and third terms are equal in calculating the rates of reactions of identical particles, such as deuterons. Despite the small factor preceding the integrals these terms may exceed the first term for sufficiently large concentrations of the particles. Let us note that according to equation (\ref{GrindEQ__33_}) $\Sigma _{ab} \sim N$, therefore the contribution of the asymptotic distribution function increases with increasing density.  In these terms the main contribution to the reaction rate is given by the different definition regions of the integrands. For example, for the second term the integration of the Maxwellian function over energy $\varepsilon _{a} $ is substantial for the values of the argument only slightly exceeding the temperature. In order for the value function - the reaction cross section was not negligible the domain of its argument should be taken in the order of hundreds of keV. Hence, the integral over $\varepsilon _{b} $ gives a significant contribution to the asymptotic region, where the function decays as a power law. Thus, for estimations it is possible to assume that the reduced energy in the center of mass $\varepsilon _{p} $ does not depend on $\varepsilon _{a} $. The integral to this argument is evaluated as:
\begin{equation}
\label{GrindEQ__54_}
\frac{2}{\sqrt{\pi \left(kT\right)^{3} } } \int _{0}^{\infty } d\varepsilon _{a} \sqrt{\varepsilon _{a} } \exp \left\{-\frac{\varepsilon _{a} }{kT} \right\}\sim 1
\end{equation}
In order to estimate the last or the fourth term in (\ref{eq:53}) we can calculate the ratio of its value to the value of the second term. Given the recent relation obtained we get:
\begin{equation}
 \frac{2\hbar e^{4} Z_{a}^{2} \Sigma _{al} }{\sqrt{m_{a} } } \int _{E_{0} }^{\infty } \frac{d\varepsilon _{a} }{\varepsilon _{a}^{7/2} } \approx \frac{2\hbar e^{4} Z_{a}^{2} \Sigma _{al} }{\sqrt{m_{a} } } \frac{{2\mathord{\left/ {\vphantom {2 5}} \right. \kern-\nulldelimiterspace} 5} }{E_{0}^{5/2} }.
\end{equation}
This ratio is obtained for the values $\Sigma _{ab} \sim N\sim 10^{23} cm^{3} $ and $E_{0} =5kT=5\cdot 10eV$. Note that the latter estimate is consistent with the relation  (\ref{GrindEQ__35_}).

\section{\label{sec:6}Reaction rate in a Lorentz gas}
Let us perform calculation of reaction rates conditioned by different energy ranges (the argument of the distribution function). The reaction frequency for particles of species $a$ and $b$ in the Lorentzian gas approximation, i.e. motionless target particles, is reduced to the following calculation:
\begin{equation}
\label{GrindEQ__56_}
\left\langle \sigma V\right\rangle =\int _{0}^{\infty } f\left(\varepsilon \right)V_{ab} \sigma \left(\varepsilon _{p} \right)\sqrt{\varepsilon } d\varepsilon
\end{equation}
Substituting the distribution function, we obtain:
\begin{eqnarray}
\left\langle \sigma V\right\rangle &=&\int _{0}^{\infty }\left(f_{0} \left(\varepsilon \right)+f_{qt} \left(\varepsilon \right)\right)V\sigma \left(\varepsilon \right)\sqrt{\varepsilon} d\varepsilon\\
\nonumber
&=&\int_{0}^{\infty }\left(\frac{2}{\sqrt{\pi } \, \left(kT\right)^{3/2}} \exp \left\{-\frac{\varepsilon }{kT} \right\}+f_{qt} \left(\varepsilon \right)\right) \sqrt{\frac{2\varepsilon _{p} }{\mu _{ab} }} S\left(0\right)\exp \left\{-\pi \sqrt{\frac{E_{G} }{\varepsilon _{p} } } \right\}\frac{\sqrt{\varepsilon}}{\varepsilon_p} d\varepsilon.
\end{eqnarray}
Calculating the terms separately, we get:
\begin{equation}
\label{GrindEQ__58_}
\left\langle \sigma V\right\rangle _{0} =\frac{4}{3} \sqrt{\frac{2kT}{\mu _{ab} } } \frac{S(0)}{kT} \tau ^{1/2} e^{-\tau }
\end{equation}
\begin{equation}
\label{GrindEQ__59_} \tau =3\left(\frac{\pi }{2} \right)^{2/3} \left(\frac{E_{G} }{kT} \right)^{1/3}
\end{equation}
The reaction rate determined by the asymptotic part of the distribution function is found from
\begin{eqnarray}
\nonumber
\left\langle\sigma V\right\rangle_{{\rm qt}}&=&\int_{0}^{\infty }f_{qt}\left(\varepsilon_{a} \right) V_{ab}\,\sigma _{f} \left(\varepsilon _{p} \right)\sqrt{\varepsilon _{a} } d\varepsilon _{a} =\\
&=&\frac{2\sqrt{2}\hbar e^{4} Z_{a}^{2}\Sigma_{al}}{\sqrt{m_a}}
\int_{0}^{\infty}\sqrt{\frac{2\varepsilon_p}{\mu _{ab}}}
S(0)\exp \left\{-\pi \sqrt{\frac{E_{G} }{\varepsilon_p}} \right\} \frac{\sqrt{\varepsilon_a}}{\varepsilon_a^4\varepsilon_p} d\varepsilon_{a}\\
\nonumber
&=&\frac{4\hbar e^{4} Z_a^2\Sigma_{al}}{\sqrt{m_{a}\mu _{ab}}}S(0)
\int_{0}^{\infty }
\exp \left\{-\pi \sqrt{\frac{E_{G} }{\varepsilon _{p} } } \right\}
\frac{\sqrt{\varepsilon_a}}{\sqrt{\varepsilon_p}\varepsilon_{a}^{4}}d\varepsilon_{a}.
\end{eqnarray}
When calculating the rate of fusion of identical particles, such as dd reaction, in the center-of-mass system for particles of equal masses and equal energies we have:
\begin{eqnarray}
\nonumber
\varepsilon _{ab} &=&\mu _{ab} \left(\frac{\varepsilon _{a} }{m_{a} } +\frac{\varepsilon _{b} }{m_b}-2\sqrt{\frac{\varepsilon_a}{m_a} \frac{\varepsilon _{b} }{m_{b} } } \cos \left(\vec{p}_{a} ,\vec{p}_{b} \right)\right)
=2\mu_{ab}\frac{\varepsilon_a}{m_a} \left(1-\cos \left(\vec{p}_{a} ,\vec{p}_{b} \right)\right)\\
&=&2\frac{m_a m_a}{m_a+m_a} \frac{\varepsilon _{a} }{m_{a} } \left(1-\cos \left(\vec{p}_{a} ,\vec{p}_{b} \right)\right)=\varepsilon_{a} \left(1-\cos \left(\vec{p}_{a} ,\vec{p}_{b} \right)\right)\approx \varepsilon_{a}.
\end{eqnarray}
Finally, we come to
\begin{eqnarray}
\left\langle \sigma  V\right\rangle _{1}&=&\frac{4 \hbar e^4 \Sigma_{al}}{\sqrt{m_a}}   S(0)\sqrt{\frac{1}{{m_a  \mathord{\left/{\vphantom{m_a 2}}\right.\kern-\nulldelimiterspace} 2} } }   \int_{0}^{\infty }
\exp \left\{-\pi \sqrt{\frac{E_{G} }{\varepsilon}} \right\}
\frac{\sqrt{\varepsilon }}{\varepsilon^4\sqrt{\varepsilon}} d\varepsilon\\
\nonumber
&=&\frac{4\sqrt{2} \hbar e^{4} Z_{a}^{2}\Sigma _{al} S(0)}{m_{a}}
\int_0^{\infty}\exp \left\{-\pi \sqrt{\frac{E_{G} }{\varepsilon } } \right\}
\frac{d\varepsilon}{\varepsilon^4}\\
\nonumber
&=&\frac{4\sqrt{2}\hbar e^{4} Z_{a}^{2} \Sigma_{al} S(0)}{m_a}\frac{2\cdot5!}{\pi ^{6} E_G^3}
=\frac{8\cdot5! \sqrt{2 } \hbar e^{4} Z_{a}^{2} \Sigma _{al} }{\pi^6 E_G^2 m_a}\frac{S(0)}{E_G}.
\end{eqnarray}
The intermediate integral used in this equation has been calculated as following:
\begin{eqnarray}
&&\int_{0}^{\infty}\exp \left\{-\pi \sqrt{\frac{E_{G} }{\varepsilon } } \right\}
\frac{d\varepsilon}{\varepsilon^4}
=2\int_{0}^{\infty}\frac{x^{8} e^{-\alpha x} }{x^{3} } dx
=2\int_{0}^{\infty}x^{5} e^{-\alpha x}dx\\
\nonumber
&&=2\frac{\partial ^{5} }{\partial \alpha ^{5} } \left(-\int _{0}^{\infty }x^{5} e^{-\alpha x} dx \right)_{\alpha =\pi \sqrt{E_{0}}}
=2\frac{\partial^5}{\partial\alpha^5}\left(\left. \frac{e^{-\alpha x} }{\alpha } \right|_{0}^{\infty } \right)_{\alpha =\pi\sqrt{E_0}} \\
\nonumber
&&
=2\frac{\partial^5}{\partial \alpha^5} \left(\frac{-1}{\alpha } \right)
=2\frac{\partial ^{4} }{\partial \alpha^4} \left(\frac{1}{\alpha^2} \right)
=2\cdot2\frac{\partial^3}{\partial \alpha^3} \left(\frac{-1}{\alpha^3} \right)=...=\frac{2\cdot5!}{\alpha^6} =\frac{2\cdot5!}{\pi ^{6}E_G^3}.
\end{eqnarray}
In this calculation we performed the change of variables:
\begin{equation}
\label{GrindEQ__64_}
x=\frac{1}{\sqrt{\varepsilon } } ,\; \; dx=-\frac{1}{2\varepsilon ^{3/2} } d\varepsilon ,\; \; d\varepsilon =-2\frac{dx}{x^{3} }
\end{equation}
In the center-of-mass for particles of identical masses but different energies, such as $\varepsilon _{a} >>\varepsilon _{b} $ we have
\begin{equation}
\label{GrindEQ__65_}
\varepsilon _{ab} =\mu _{ab} \left(\frac{\varepsilon _{a} }{m_{a} } +\frac{\varepsilon _{b} }{m_{b} } -2\sqrt{\frac{\varepsilon _{a} }{m_{a} } \frac{\varepsilon _{b} }{m_{b} } } \cos \left(\vec{p}_{a} ,\vec{p}_{b} \right)\right)\approx \mu _{ab} \frac{\varepsilon _{a} }{m_{a} }  \end{equation}
Finally, for such a case we get
\begin{eqnarray}
\nonumber
\left\langle\sigma V\right\rangle_{2}&=&\frac{4\hbar e^4\Sigma_l}{\sqrt{m_a}}   S(0)\sqrt{\frac{1}{\mu_{ab}}}
\int_{0}^{\infty}\exp\left\{-\pi\sqrt{\frac{E_G}{\varepsilon}}\right\}
\frac{\sqrt{\varepsilon}d\varepsilon}{\varepsilon^4\sqrt{\varepsilon}}\left(\frac{\mu_{ab}}{m_a} \right)^{5/2}\\
&=&\frac{4\hbar e^{4}\mu_{ab}^2 Z_a^2\Sigma_l S(0)}{m_a^3} \int_0^{\infty}
\exp \left\{-\pi\sqrt{\frac{E_G}{\varepsilon } } \right\}
\frac{d\varepsilon}{\varepsilon ^4}
\\
\nonumber
&=&\frac{4\hbar e^4\mu_{ab}^2 Z_a^2 \Sigma_l S(0)}{m_a^3}   \frac{2\cdot5!}{\pi^6 E_G^3}=\frac{8\cdot5!}{\pi^6} \frac{S(0)}{E_G}\frac{\hbar e^4}{E_G^2}\frac{\mu_{ab}^2 Z_a^2\Sigma_l }{m_a^3}.
\end{eqnarray}
Given the equal masses of particles we get
\begin{equation}
\label{GrindEQ__67_}
\left\langle \sigma V\right\rangle _{2} =\frac{8\cdot 5!}{\pi ^{6} } \frac{S(0)}{E_{G} } \frac{\hbar e^{4} }{E_{G}^{2} } \frac{m_{a}^{2} Z_{a}^{2} \Sigma _{l} }{4m_{a}^{3} } =\frac{2\cdot 5!}{\pi ^{6} } \frac{S(0)}{E_{G} } \frac{\hbar e^{4} }{E_{G}^{2} } \frac{Z_{a}^{2} \Sigma _{l} }{m_{a} }
\end{equation}

\section{\label{sec:7}Analytical estimates for reaction rates for model distributions with quantum effects}
For the integrals in the second and third terms in (\ref{eq:53}), different ranges of $\varepsilon _{a} $ and $\varepsilon _{b} $ are essential. For the second term the integrand is close to $0$ if the energy is greater than the temperature, i.e. outside the interval $0<\varepsilon _{a} <3kT$  For variable $\varepsilon _{b} $ this integral area is significantly wider. In the third term variables $\varepsilon _{a} $ and $\varepsilon _{b} $ interchange.

Let us now consider the second term. We assume $\varepsilon _{a} <<\varepsilon _{b} $ taking into account the significant range of energies of different particles. Then the energy in the center of mass can be represented by
\begin{equation}
\label{GrindEQ__68_}
\varepsilon _{p} =\mu _{ab} \left(\frac{\varepsilon _{a} }{m_{a} } +\frac{\varepsilon _{b} }{m_{b} } -2\sqrt{\frac{\varepsilon _{a} }{m_{a} } \frac{\varepsilon _{b} }{m_{b} } } \cos \left(\vec{p}_{a} ,\vec{p}_{b} \right)\right)\approx \mu _{ab} \frac{\varepsilon _{b} }{m_{b} }  \end{equation}
If we substitute this expression into the second term in (\ref{eq:53}) it turns out that the rate and the cross section do not depend on the angle between the velocity vector and the integral is reduced to
\begin{eqnarray}
\nonumber
\left\langle\sigma V\right\rangle_{b}&=&\frac{2}{\sqrt{\pi}\left(kT\right)^{3/2}} \frac{\sqrt{2}\hbar e^{4}Z_{b}^{\,2}\Sigma_{bl}}{\sqrt{m_b}} \int_{0}^{\infty}d\varepsilon_{a}\sqrt{\varepsilon_a}\exp\left\{-\frac{\varepsilon_a}{kT} \right\}\int_{0}^{\infty}\frac{d\varepsilon_b}{\varepsilon _{b}^{7/2}}     V_{ab} \left(\varepsilon_p\right)\sigma \left(\varepsilon_p\right)2\\
&=&\frac{2\sqrt{2}\hbar e^{4} Z_{b}^{2} \Sigma_{bl}}{\sqrt{m_b}}\int_{0}^{\infty}
\frac{d\varepsilon_b}{\varepsilon_b^{7/2}}V_{ab}\left(\varepsilon_p\right)\sigma \left(\varepsilon_p\right)\\
\nonumber
&=&\frac{2\sqrt{2} \hbar   e^{4} Z_{b}^{2} \Sigma _{bl}}{\sqrt{m_b}} \left(\frac{\mu_{ab} }{m_b} \right)^{5/2} \int _{0}^{\infty }\frac{d\varepsilon_p}{\varepsilon _{p}^{7/2}}  V_{ab} \left(\varepsilon_p \right)\sigma \left(\varepsilon_p\right)\\
\nonumber
&=&\frac{2\sqrt{2} \hbar   e^{4} Z_{b}^{2} \Sigma _{bl} }{\sqrt{m_{b} } } \left(\frac{\mu _{ab} }{m_{b} } \right)^{5/2} \int _{0}^{\infty }\frac{d\varepsilon _{p} }{\varepsilon _{p}^{7/2}}    \sqrt{\frac{2\varepsilon _{p} }{\mu _{ab} } }\frac{S(0)}{\varepsilon_p}
\exp\left\{-\pi\sqrt{\frac{E_G}{\varepsilon_p}}\right\}\\
\nonumber
&=&\frac{4\hbar e^{4} Z_{b}^{2} \Sigma _{bl} }{m_{b} } S\left(0\right)\left(\frac{\mu _{ab} }{m_{b} }\right)^{2}\int _{0}^{\infty }\frac{d\varepsilon _{p} }{\varepsilon _{p}^{7/2}\sqrt{\varepsilon_p}}
\exp\left\{-\pi \sqrt{\frac{E_G}{\varepsilon_p } } \right\}\\
\nonumber
&=&\frac{8\cdot5!}{\pi^6}\frac{S(0)}{E_G}\frac{\hbar e^{4}}{E_{G}^{2}} \left(\frac{\mu_{ab}}{m_{b} }\right)^2\frac{Z_{b}^{2}\Sigma_{bl}}{m_{b}}.
\end{eqnarray}
We can practically perform similar transformations as earlier but here we have
\begin{equation}
\label{GrindEQ__70_}
Z_{b}^{2} \Sigma _{bl} =Z_{b}^{2} m_{b}^{2} \sum _{l} \frac{N_{l} Z_{l}^{2} }{\mu _{bl}^{2} }  \end{equation}
For the third tern in (\ref{eq:53}) we get
\begin{equation}
\label{GrindEQ__71_}
\left\langle \sigma V\right\rangle _{a} =\frac{8\cdot 5!}{\pi ^{6} } \frac{S\left(0\right)}{E_{G} } \frac{\hbar e^{4} }{E_{G}^{2} } \left(\frac{\mu _{ab} }{m_{a} } \right)^{2} \frac{Z_{a}^{2} \Sigma _{al} }{m_{a} }  \end{equation}
where
\begin{equation}
\label{GrindEQ__72_}
Z_{a}^{2} \Sigma _{al} =Z_{a}^{2} m_{a}^{2} \sum _{l} \frac{N_{l} Z_{l}^{2} }{\mu _{al}^{2} }  \end{equation}
Thus, the reaction rate corrected with the quantum tail contribution is equal to
\begin{equation}
\label{GrindEQ__73_} \left\langle \sigma V\right\rangle _{quant} =\left\langle \sigma V\right\rangle _{a} +\left\langle \sigma V\right\rangle _{b} =\frac{8\cdot 5!}{\pi ^{6} } \frac{S(0)}{E_{G} } \frac{\hbar e^{4} \mu _{ab}^{2} }{E_{G}^{2} } \left(\frac{Z_{a}^{2} }{m_{a} } \sum _{l} \frac{N_{l} Z_{l}^{2} }{\mu _{al}^{2} } +\frac{Z_{b}^{2} }{m_{b} } \sum _{l} \frac{N_{l} Z_{l}^{2} }{\mu _{bl}^{2} } \right)
\end{equation}
In the sums of the last formula the summation index \textit{l} should stand for all particles of the medium except for species \textit{a} or \textit{b} respectively. It follows from (\ref{GrindEQ__40_}) and the following comments. Because of the small correlation function of identical particles, the quantum corrections corresponding to such scattering are also small.

Let us note that the reaction rate constant determined for the tails of the distribution function and represented as (\ref{GrindEQ__73_}) does not depend explicitly on temperature. As noted above, this is due to the fact that in the asymptotic region the tails of the distribution function do not depend on temperature. At the same time it should be noted that temperature defines the ionic composition of the plasma. Therefore, the temperature dependence of the distribution function is extremely important and is determined by the terms in brackets in (\ref{GrindEQ__73_}). As it is shown below, the definition of the ionic composition of plasma is essential for temperature range of 5-10 eV. It must be noted, that correct definition  of  the distribution for ionic composition is very complicated problem in nonideal plasmas, so the measurement of neutron yield gives an instrument to have information on equation of state in such plasmas.

Thus, we have given proof of the validity of the approximate approach, which consists in averaging the cross sections found using the quantum corrections to the momentum distribution functions of the reacting particles. The accuracy of this approach is discussed further below where the results of calculations are shown for the full formula (\ref{eq:2}), simplified equation (\ref{eq:17}), and approximate analytical estimations (\ref{GrindEQ__58_}) and (\ref{GrindEQ__71_}). This eliminates the issues raised in the paper \cite{Zubarev2008} on the validity of such a method.

From (\ref{GrindEQ__73_}) we can estimate the rate constant for fusion of deuterons taking 49.6 keV barn for an astrophysical factor $S(0)$ and to 99.9 keV for the Gammov energy $E_G$. Calculating the scattering cross section at $E_G$ as well as estimating the velocity of the particles appearing in the reaction rate, we find that the collisional width of the gamma at the density of scattering particles of the order of $3\times10^{23}$ cm$^{-3}$ appears to be about 10-6 eV and the reaction rate constant turns out to be of the order of 10-28 cm$^3$ c$^{-1}$.

In a coupled plasma the Maxwellian contribution to the reaction rate is specified to make an account for screening of the Coulomb potential \cite{Salpeter1954, Chugunov2007, Brown1997, Bahcall2002}. The corresponding generalized expression is obtained, for example, by replacing the exponential Sommerfeld factor in equation (\ref{eq:7}) with the semiclassical tunneling probability through the screened potential barrier:
\begin{equation}
\label{GrindEQ__74_}
\exp \left\{-2\pi \eta \right\}\to \exp \left\{-\frac{2\sqrt{2\mu _{ab} } }{\hbar } \int _{r_{n} }^{r_{l} } \sqrt{\frac{Z_{a} Z_{b} e^{2} }{r} -H(r)-E_{p} } dr\right\}
\end{equation}
where $r_{n} $ and $r_{l} $ are classic stopping points and $Z_{a} Z_{b} e^{2} /r-H\left(r\right)$ is the interaction potential with allowance for screening effects. It may be noted that the progress in the theory of the subbarrier tunneling, connected with the problem of heavy ion fusion, was analyzed in \cite{Balantekin1998}.

Great interest was aroused by the experimental studies of reactions involving light nuclides such as isotopes of hydrogen, He, and Li. They have shown the exponential enhancement of the nuclear reactions cross-sections at low energies \cite{Aliotta2000, Strieder2001, Raiola2002, Luo2003,  Czerski2006, Huke2007}. The beam of the light nuclides ions was focused on the metallic target. The Coulomb screening of electrons had been considered in this case. The electrons respond by accumulating around the positive charge and therefore partially screen out its positive Coulomb potential. Although the experiments have proved the significance of electron screening, a theoretical explanation is still far from satisfactory. The high screening potential value arises from the environment of the light nuclides in the metallic matrix, but a quantitative explanation is missing. The screening effect was far beyond the expected value.

It must be noted that for the quantum correction the screening effects can be ignored because the screening energy is of the order of $H(0)$, and is therefore small compared to the Gamow energy $E_G$, which determines the ``tail'' contribution to the value of reaction rates.

\section{\label{sec:8}Modeling of fusion reactions}
Nuclear fusion reactions occur at an appreciable rate in conditions of hot plasma. This is primarily the plasma in stars; in particular, it is the plasma in the solar interior. From the models considered in this paper, the quantum effects can also predict the increase in the rate of reactions in plasmas of moderate temperatures but high densities. By monitoring the synthesis reaction, this increase might be observed for suitable parameters.  In order to determine these parameters, we calculated the fusion rates for different conditions. It is also interesting to do some revision of processes in the interiors of stars, to re-evaluate the contribution of various processes to the release of fusion energy, and ultimately to the evolution of stars.

Table II shows the constants, which were calculated for the fusion reaction $p+p\to D+e^{+} +\nu $ at three points along the solar trajectory, from the central part to the periphery. In the table we show the constants calculated numerically:  $\langle V_{ab}\rangle\_3$, $\langle V_{ab}\rangle\_5$, $\langle V_{ab}\rangle\_10$ as well as analytical estimations of constant for the reaction, which were calculated using equation (58) (Classic rate), the quantum correction of constant for the reaction, which were calculated with equation (73) (Quantum correction), and the sum of these two values (Full rate). As one can see the calculations performed for different models are in a satisfactory agreement with each other.
\begin{table}
  \centering
\begin{tabular}{|c|c|c|c|}
\hline
&$\,\bf{N=3.31   10^{25}cm^{-3}}\,$ & $\,\bf{N=9.575431   10^{24}cm^{-3}}\,$& $\,\bf{N=7.405   10^{22}cm^{-3}}\,$\\
&$\bf 1.3362\quad KeV$ & $\bf 0.6892\quad KeV$ & $\bf 0.1828\quad KeV$ \\
\hline
$\textrm{Classic rate}$ & $2.98 10^{-45}$ & $6.13 10^{-47}$ & $0.79 10^{-51}$ \\
\hline
$\textrm{Quantum correction}$   & $2.77 10^{-49}$  & $0.8\cdot10^{-49}$  & $0.62\cdot10^{-51}$ \\
\hline
$\textrm{Full rate (analitic)}$ & $ 2.98\cdot10^{-45}$ & $6.13\cdot10^{-47}$ & $1.41\cdot10^{-51}$ \\
\hline
$\left\langle\sigma V_{ab}  \right\rangle \_3$  & $3.27\cdot10^{-45}$ & $7.45\cdot10^{-47}$ & $1.28\cdot10^{-51}$ \\
\hline
$\left\langle\sigma V_{ab}  \right\rangle \_5$  & $3.08\cdot10^{-45}$ & $6.78\cdot10^{-47}$ & $1.43\cdot10^{-51}$ \\
\hline
$\left\langle\sigma V_{ab}  \right\rangle \_10$ & $3.35\cdot10^{-45}$ & $6.48\cdot10^{-47}$ & $1.61\cdot10^{-51}$ \\
\hline
\end{tabular}
\caption{\label{T:2}}
\end{table}

Another conclusion to be drawn from the results presented in this table is that the influence of the quantum effects is notable for this reaction only in a distant region from the Sun center. But in this region the fusion rate is much lower than in the central region, so the influence of the quantum corrections brings no effect on the energy balance for this reaction and synthesis of deuterons in the solar interior. Our simulations were carried out using the data on the astrophysical factor from \cite{Adelberger1998}.

An example of the reaction with the rate greatly influenced by the quantum corrections is the reaction of the hydrogen cycle: $^{3} He+^{3} He\to 2p+^{4} He$. Table III shows the rate constants calculated for the same points of the solar trajectory. In calculations of these rates the value of the astrophysical factor was taken from \cite{CF88}.
\begin{table}
    \centering
\begin{tabular}{|c|c|c|c|}
\hline
& $\,\bf{N=3.31\cdot10^{25}cm^{-3}\,}$ & $\,\bf{N=9.575431\cdot10^{24}cm^{-3}}\,$ & $\,\bf{N=7.405\cdot10^{22}cm^{-3}}$\,\\
& $  \bf{1.3362\quad KeV} $ & $  \bf{0.6892\quad KeV} $ &  $ \bf{0.1828\quad KeV}$ \\
\hline
$\textrm{Classic rate}$ & $ 6.84 \cdot10^{-34} $ & $ 5.65  \cdot10^{-39} $ & $ 1.86  \cdot10^{-53}$ \\
\hline
$\textrm{Quantum correction}$ & $ 2.34  \cdot10^{-27} $ & $ 5.37  \cdot10^{-28} $ & $ 4.14  \cdot10^{-30}$ \\
\hline
$\textrm{Full rate (analitic)}$ & $ 2.34  \cdot10^{-27} $ & $ 5.37  \cdot10^{-28} $ & $ 4.14  \cdot10^{-30}$ \\
\hline
$\left\langle\sigma V_{ab}  \right\rangle\_3 $ & $ 2.38  \cdot10^{-27} $ & $ 5.42  \cdot10^{-28} $ & $ 4.15  \cdot10^{-30}$ \\
\hline
$\left\langle\sigma V_{ab}  \right\rangle\_5 $ & $ 1.34  \cdot10^{-27} $ & $ 3.03  \cdot10^{-28} $ & $ 2.24  \cdot10^{-30}$ \\
\hline
\end{tabular}
\caption{\label{T:3}}
\end{table}

Similarly, it could be shown that the rates of many fusion reactions, which occur in the solar interior, such as$p+^{7} Be$, $^{3} He+^{4} He$, etc., as well as of reactions such as $C+C$ in the depths of supernovae, become much larger than their classic values if the quantum effects are taken into account. These results require further consideration and beyond the scope of this work.

However, these predictions, in principle, might be verified in laboratory experiments using a dense plasma with moderate temperatures around one electron volt and above. This can be achieved in explosive experiments like the ones, for example, that study the equation of state of strongly coupled plasma \cite{Fortov2003, Grishechkin2004, Fortov2007}.  If we take deuterium, compressed to a pressure of the order of megabars, we can make the following prediction: in pure deuterium, in which there is a noticeable degree of dissociation and ionization due to pressure ionization, the reaction rate will be very small in these conditions because of the factor $g(0)$ in (\ref{GrindEQ__40_}), which in turn is small in a strongly coupled plasma. If deuterium is diluted with an extraneous gas, such as a different isotope of hydrogen, helium, etc., then there will be terms in the $D+D$ reaction rate due to scattering on a buffer gas, which can lead to an observable neutron yield of about $10^8$ to $10^{10}$ neutrons per pulse of about 1 microsecond length. The result strongly depends on the ion composition and possibility of the plasma phase transition in strongly coupled plasmas.

As another example, we present calculations for a mixture of deuterium and xenon, which can create a shock wave with a speed exceeding $5$ km/c and create a plasma with temperature of $5-10$ eV in the reflected wave. Calculations of the plasma parameters for these conditions were kindly provided by V.K. Gryaznov.

Table \ref{T:4} shows the results of the calculation of constant fusion reaction: $D+D\to 3He+n$ for different spatial points of the plasma in conditions of shock compression of a mixture of deuterium and xenon. The gas mixture was $D:Xe=50:50$ in volume ratio. The initial pressure was 25 bars. The table shows the equilibria concentration of various components of plasma for these conditions: $\mathrm{N(D^+)}$, $\mathrm{N(Xe^+)}$ etc, reaction constants $\langle\sigma V\rangle$ and the rates of the reaction $\langle\sigma V\rangle\mathrm{N}(\mathrm{D}^+)^2$. The latter ones have been obtained analytically by taking into account the quantum corrections.
\begin{table}
    \centering
\begin{tabular}{|c|c|c|c|c|}
\hline
$\mathbf{T,\,eV }$ & $\quad9.27\quad $ & $\quad9.88\quad $ & $\quad1.06\quad $ & $\quad1.22\quad$ \\
\hline
$\mathbf{N(D^+)}$ & $ 5.45\cdot10^{21} $ & $ 5.54\cdot10^{21} $ & $ 5.63\cdot10^{21} $ & $ 5.74  \cdot10^{21}$ \\
\hline
$\mathbf{N(Xe^+)}$ & $ 6.35 \cdot10^{20} $ & $ 2.79\cdot10^{20} $ & $ 1.46  \cdot10^{20} $ & $ 6.24  \cdot10^{19}$ \\
\hline
$\mathbf{N(Xe^{+2})}$ & $ 5.53  \cdot10^{21} $ & $ 5.19 \cdot10^{21} $ & $ 4.61  \cdot10^{21} $ & $ 3.31 \cdot10^{21}$ \\
\hline
$\mathbf{N(Xe^{+3})}$ & $ 2.26  \cdot10^{21} $ & $ 2.68  \cdot10^{21} $ & $ 3.08  \cdot10^{21} $ & $ 3.76  \cdot10^{21}$ \\
\hline
$\mathbf{N(Xe^{+4})}$ & $ 8.82  \cdot10^{19} $ & $ 1.40  \cdot10^{20} $ & $ 2.19 \cdot10^{20} $ & $ 4.79 \cdot10^{20}$ \\
\hline
$\langle\sigma V\rangle$ & $ 3.5  \cdot10^{-29} $ & $ 3.68 \cdot10^{-29} $ & $ 3.85 \cdot10^{-29} $ & $ 4.13 \cdot10^{-29}$ \\
\hline
$\langle\sigma V\rangle\mathbf{N}(\mathbf{D}^+)^2$ & $ 1.04\cdot10^{15} $ & $ 1.13\cdot10^{15} $ & $ 1.22 \cdot10^{15} $ & $ 1.36\cdot10^{15}$ \\
\hline
\end{tabular}
\caption{\label{T:4}}
\end{table}

As this table shows, the temperature of the plasma in such conditions is more than 2 orders of magnitude lower than the temperature of the Sun plasma. The plasma density is sufficiently high. Under the conditions of shock compression of Xenon is 2 to 3-times ionized as a result of pressure ionization \cite{CF88}. The calculations used data on the astrophysical factor of the reaction of \cite{Huba:book}.

For the conditions of the shock experiments it would be important to estimate the relaxation time of the distribution function. In \cite{Starostin2005UFN, Galitsky1966} it was suggested, that the only binary elastic scattering must be taken into account to study relaxation of the nonequilibrium distribution function. It is the Coulomb type collision for the plasma in a shock wave. The cross section for these collisions decreases with increasing energy. So the elastic scattering frequency for the kinetic energy is of the order of the Gamow energy. When the fusion reaction is realized distinctly, it may be estimated as:
\begin{equation}
\label{GrindEQ__75_}
\left\langle \sigma _{t} V\right\rangle \approx \frac{2\pi e^{4} Z_{a}^{2} Z_{b}^{2} }{E_{G}^{2} } \sqrt{\frac{2E_{G} }{\mu _{ab} } }
\end{equation}
Then the upper level of the relaxation time is $\tau \approx (N\left\langle \sigma _{t} V\right\rangle )^{-1} $ with the data from Table \ref{T:4}. In this case it may be estimated as about 1 ns to 10 ns. In \cite{Krook1976} the period of relaxation of the nonequilibrium distribution function was calculated more precisely. The model, used in this paper, permits an analysis of the relaxation process for all velocities, including formation of the distribution function tail. In this paper for the various dependencies of the elastic scattering cross section on energy it was obtained that the time of the equilibrium setting is larger, up to factor 6, than that of the elastic collisions. This estimation was obtained for the energy equal to the kinetic temperature. The calculated value of relaxation is less than 1 ns.

Thus, for estimating the fusion reaction, it can be seen that the collisions in the plasma of the shock experiments are effective in bringing the distribution function to equilibrium during the hydrodynamical process with the characteristic time of $<1$ $\mu$s.

The fusion rate constants for the above conditions calculated with no account taken for the quantum effects are by about 20 orders of magnitude smaller than the values listed in the table. The reaction rate for different points of the plasma vary by ${\rm \sim }30\% $, as it is seen from the last line of the table. The lifetime of such a plasma is of ${\rm \sim }1$ $\mu $s, which gives the neutron yield of ${\rm \sim }1.2\times10^{9} $ cm$^{-3}$.

\section{\label{sec:9}Conclusions}
In this paper we analyzed the influence of quantum effects on the rate of fusion reactions. As a result of frequent collisions of particles in a dense plasma there disappears the complete correspondence between the total and kinetic energy of the particle, the generalized distribution function, thus, depends on both the total and the kinetic energy of the particle. The momentum distribution function has the power dependence on the kinetic energy in the asymptotic region. Carrying out the averaging over the distribution function to calculate, for example, the fusion reaction rate leads to a notable increase in reaction rate as compared to the calculations using the Maxwellian distribution function. We created numerical models for various conditions of the reactions and carried out calculations in a wide range of plasma parameters. The use of the approximate analytical estimates obtained under the averaging procedure has been validated.

In addition, the ranges of parameters where the most pronounced quantum effects are expected are shown. Two experiments are identified that might demonstrate quantum effects:  one, by comparing DD reaction rates in densely compressed cold deuterium plasma with or without a buffer gas; and, two, by arranging for a shock wave in mixtures of Deuterium and Xenon.

The presence of quantum tails is also evident in the vibrational kinetics of low-temperature plasma. It increases the rate of V-T relaxation, as shown in \cite{Starostin2005UFN}. This result of the theory is in a very good agreement with the experimental data under normal conditions. The theory also predicts the reduction of the induction time for ignition of hydrogen-oxygen and hydrogen-air mixtures at pressures above 5 atm and at temperatures below 1000 K.

\begin{acknowledgments}
The authors wish to express their sincere appreciation and gratitude to V.E. Fortov, V.B. Mintsev, V.K. Gryaznov, I.V. Lomonosov, N. Shilkin, S.V. Ayukov, V.A. Baturin, A.B. Gorshkov and S. Taova for fruitful discussions and valuable advice. This work was supported by the ISTC project No. 3755 and partially by grant NSh-3239.2010.2.
\end{acknowledgments}

\appendix
\section{Monte Carlo integrations}
Calculation of the approximate value of the integral using the Monte Carlo method is one of the few methods for calculating the quadrature in this problem, given the high multiplicity of integral. Optimization of the calculations in the framework of this method is the choice of determining the probability distribution of sites of integration of the quadrature formula. On one hand it minimizes the dispersion of the mean value while on the other hand it minimizes the possibility of a fairly simple and efficient simulation of random vectors, which determine the quadrature grid of the numerical integration.

Minimum dispersion in the calculation of the integral is achieved when selecting a random distribution function of several variables, proportional to the integrand \cite{Ermakov:book}. Given this condition, the probability density of the random vector in a multidimensional space was defined as the product of functions, which were probability densities for individual components of the random vector. In the Monte-Carlo method the integration variables are the very random variables. Functions that are selected as probability densities constitute a significant part of the integrand function provided in the multiplicative form. For example, we consider in detail the computation of the integral, which reduces the calculation of the reaction rate for the non-degenerate distribution function (\ref{eq:17}):
\begin{eqnarray}
\label{eq:76}
S_{ab} &=&\int_{0}^{\infty}dE_{a}
\int_{0}^{\infty}d\varepsilon_a \sqrt{\varepsilon_a}
\int_{0}^{\infty }dE_{b} \int _{0}^{\infty}d\varepsilon_b\sqrt{\varepsilon_b}
\int_{-1}^{1}d\cos (\vec{p}_{a},\vec{p}_{b})\\
\nonumber
&\times& n(E_{a})\delta\gamma_a\left(E_{a}-\varepsilon_a,\varepsilon_a\right)n(E_{b})
\delta\gamma_b\left(E_b-\varepsilon_b,\varepsilon_{b}\right) V_{ab}\,\sigma\left(\varepsilon_p \right).
\end{eqnarray}
In accordance with the foregoing notes, in integration over the variables $E_{a} $, $E_{b} $ we use $n\left(E_{a} \right)$, $n\left(E_{b} \right)$ as probability density functions, and over the variables $\varepsilon _{a} $, $\varepsilon _{b} $ - functions of the spectral particle characteristic $\delta \gamma _{a} (E_{a} -\varepsilon _{a} ,\varepsilon _{a} )$ and $\delta \gamma _{b} (E_{b} -\varepsilon _{b} ,\varepsilon _{b} )$.

In accordance with the technique used in this paper we will decompose the factors into terms corresponding to the probability density and a factor, the average value of which is to be calculated. The following expressions describe for the probabilities of random variables, i.e. the variables of integration (\ref{eq:76}) with the selected probability density:
\begin{equation}
\label{GrindEQ__77_} S_{3a} \left(E_{a} \right)=\alpha _{3a} \int _{0}^{E_{a} } dE_{a} n\left(E_{a} \right),0\le E_{a} <\infty
\end{equation}
\begin{equation}
\label{GrindEQ__78_}
S_{2a} \left(\varepsilon _{a} \right)=\alpha _{2a} \int _{0}^{\varepsilon _{a} } d\varepsilon _{a} \delta \gamma _{a} \left(E_{a} -\varepsilon _{a} ,\varepsilon _{a} \right),0\le \varepsilon _{a} <\infty
\end{equation}
\begin{equation}
\label{GrindEQ__79_}
S_{3b} \left(E_{b} \right)=\alpha _{3b} \int _{0}^{E_{b} } dE_{b} n\left(E_{b} \right),\quad \quad 0\le E_{b} <\infty
\end{equation}
\begin{equation}
\label{GrindEQ__80_}
S_{2b} \left(\varepsilon _{b} \right)=\alpha _{2b} \int _{0}^{\varepsilon _{b} } d\varepsilon _{b} \delta \gamma _{b} \left(E_{b} -\varepsilon _{b} ,\varepsilon _{b} \right),0\le \varepsilon _{b} <\infty  \end{equation}
\begin{equation}
\label{GrindEQ__81_}
S_{1} \left(x\right)=\alpha _{1a} \int _{-1}^{x} dx,-1\le x<1.
\end{equation}
Here $S_{\psi } \left(z\right)$ is the probability of a random variable to be in the range with the upper limit of $z$, the factors $\alpha _{\psi } $ are the normalizing factors that ensure the implementation of the normalization condition for the probability of the variable denoted by the index $\psi $.

Distribution of random variables with a given probability density is performed using the standard methods. It is necessary to find a solution for the system of nonlinear equations:
\begin{equation}
\label{GrindEQ__82_}
S_{\psi } \left(z\right)=u_{\psi },
\end{equation}
where index $\psi$ runs over all the values corresponding to different variables of integration, $u_{\psi } $ are random variables uniformly distributed over the range [0,1]. Variables $E_{a} $, $E_{b} $, $x$ can be determined from one of the equations in system (\ref{GrindEQ__77_}), (\ref{GrindEQ__79_}), and (\ref{GrindEQ__81_}). The selected probability densities for variables $\varepsilon _{a} $, $\varepsilon _{b} $ contain other integration variables in addition to their ``own'' variables. Therefore, during simulation of these random variables, we have to solve the system of two nonlinear equations (\ref{GrindEQ__78_}) and (\ref{GrindEQ__80_}) applying the method of iterations.

The value of integral (\ref{eq:76}) is found as the average value of the expression:
\begin{equation}
\label{GrindEQ__83_}
S_{ab} =\frac{1}{N} \sum _{1}^{N} \sqrt{\varepsilon _{a} } \sqrt{\varepsilon _{b} } V_{ab} {\kern 1pt} \sigma (\varepsilon _{p} ),
\end{equation}
with variables $\varepsilon _{a} $ and $\varepsilon _{b} $ as well as values of $V_{ab} $ and $\sigma (\varepsilon _{p} )$ contained in this sum determined by the above-described drawing of random variables.

\bibliography{Starostin}

\end{document}